# Training Generative Adversarial Networks for Optical Property Mapping using Synthetic Image Data

**A. O**SMAN [1,2]**, J. C**ROWLEY [1,3] **A**ND **G.S.D G**ORDON [1,4]

[1]*Optics and Photonics Group, Faculty of Engineering, The University of Nottingham, Nottingham, United Kingdom*
[2]*ahmedosman30013@gmail.com*
[3]*jane.crowley@nottingham.ac.uk*
[4]*george.gordon@nottingham.ac.uk*

**Abstract**: We demonstrate the training of a Generative Adversarial Network (GAN) for prediction of optical property maps (scattering and absorption) using spatial frequency domain imaging (SFDI) image data sets generated synthetically with free open-source 3D modelling and rendering software, Blender. The flexibility of Blender is exploited to simulate 5 models with real-life relevance to clinical SFDI of diseased tissue: flat samples containing a single material, flat samples containing 2 materials, flat samples containing 3 materials, flat samples with spheroidal tumours and cylindrical samples with spheroidal tumours. The last case is particularly relevant as it represents wide-field imaging inside a tubular organ e.g. the gastro-intestinal tract. In all 5 scenarios we show the GAN provides accurate reconstruction of optical properties from single SFDI images with mean normalised error ranging from 1.0-1.2% for absorption and 1.1%-1.2% for scattering, resulting in visually improved contrast for tumour spheroid structures. This compares favourably with the ~10% absorption error and ~10% scattering error achieved using GANs on experimental SFDI data. Next, we perform a bi-directional cross-validation of our synthetically-trained GAN, retrained with 90% synthetic and 10% experimental data to encourage domain transfer, with a GAN trained fully on experimental data and observe visually accurate results with error of 6.3%-10.3% for absorption and 6.6%-11.9% for scattering. Our synthetically trained GAN is therefore highly relevant to real experimental samples but provides the significant added benefits of large training datasets, perfect ground-truths and the ability to test realistic imaging geometries, e.g. inside cylinders, for which no conventional single-shot demodulation algorithms exist. In the future, we expect that the application of techniques such as domain adaptation or training on hybrid real-synthetic datasets will create a powerful tool for fast, accurate production of optical property maps for real clinical imaging systems.

## 1. Introduction

Spatial Frequency Domain Imaging (SFDI), a non-contact, depth-varying and wide-field optical imaging technique, has proven especially useful in a range of biomedical imaging applications [1]. SFDI utilises spatially modulated light patterns of visible or near-infrared light to extract optical properties such as absorption, $\mu_a$, and reduced scattering, $\mu_s$', of biological tissue [2], making it possible to produce maps of relevant biomedical indicators such as blood oxygenation, hemodynamics and structural properties of tissue. All that is required to implement this imaging technique is a consumer-grade camera and a projector. The relative simplicity of this set-up makes it suitable for low-cost and miniaturised implementations, suitable for future clinical deployment, e.g. capsule endoscopes [3]. To date, SFDI has been shown to be a highly effective method of quantifying burn severity, monitoring vascular occlusions, imaging drug delivery to the brain and detecting tumour margins in oncology [4-6].

Despite its successes in a variety of clinical applications, SFDI still presents many challenges that limit the adoption of the technology in clinical practice. One particular challenge of conventional SFDI is that it requires at least 3 images per wavelength (3 different spatial phases) to create a single optical property map, which makes real-time display of optical property maps difficult. However, recent techniques have demonstrated the use of halftone patterns projected using digital micromirror devices to achieve kilohertz projection rates which can significantly speed this up and also allow use of multiple wavelengths and spatial frequencies [7, 8].

Although single snapshot imaging of optical properties (SSOP) has reduced the number of images, the method causes image artefacts to arise and requires frequency filtering [9], sacrificing quality and fine detail of the resulting optical property maps

Recently, several approaches have been proposed that use convolutional neural networks to improve image reconstruction from single images compared to SSOP. Chen et al. recently recasting production of optical property maps as a deep learning problem, a conditional generative adversarial network (cGAN) was trained to produce physically correct optical property maps from raw SFDI images of diseased and healthy tissue. Coined GANPOP (Generative Adversarial Network Prediction of Optical Properties), this cGAN has been proven to estimate the absorption and reduced scattering coefficients 60% more accurately than SSOP for single structured-light input images in human gastrointestinal specimens [10]. When trained on structured-illumination images of human specimens and homogeneous phantoms, then tested on in vivo and ex vivo swine tissue, the GANPOP performed 46% better than SSOP at predicting the optical properties [10]. Aguénounon et al. developed a deep convolutional neural network with a U-Net architecture that can accurately reconstruct AC and DC demodulated components from a single SFDI image. This produced errors typically <10% and was able to operate in real time (18ms) [11]. The nonlinear mapping between such demodulated images and absorption and scattering coefficients has also been approximated using random-forest algorithms [12] and deep learning [13].

Scattering and absorption coefficients often represent an intermediate quantity, with some other clinically relevant property, such as tissue oxygenation, being the final goal. To that end, another cGAN architecture, named OxyGAN, was recently demonstrated to produce maps of tissue oxygenation over a widefield of view from raw SFDI images. This OxyGAN showed a 96.5% accuracy when tested on human feet and when applied to samples not seen previously in the training data such as human hands and pig colons, it achieved an accuracy of 93%. Additionally, this model is 10 times faster and 24.7% more accurate than a hybrid model that combined the GANPOP with a physical model, thus enabling real-time, tissue oxygenation mapping at a rate of 25Hz [14]. GANs have also recently been applied to exploit optical properties to create maps of healthy and malignant tissue in breast cancer [15] and more generally have been applied for medical image segmentation, registration, reconstruction and disease detection [16].

Provided the training dataset spans the target domain, GANs have the potential to produce accurate optical property maps in real-time, which may in future be used to diagnose diseases more accurately or guide surgical procedures. However, there are two key challenges that may limit application to real-world problems.

The first challenge is that in some instances it is not feasible to produce experimentally-acquired datasets that are sufficiently large and diverse to span this domain, particularly if it would require deploying imaging systems in environments where they are not yet approved for use (e.g. inside the gastrointestinal tract of human subjects).

The second challenge is that conventional methods require a carefully controlled light projection, necessitating occasional re-calibration which can be difficult to achieve in clinical settings. The projected patterns used in previous systems use a plane wave basis and assume orthographic (or telecentric) projection and imaging onto samples [17, 18]. This is suitable for flat structures, e.g. skin, with minor variations such as small lumps, but is unsuitable for many clinically relevant geometries, in particular imaging inside lumen such as the gastrointestinal tract or blood vessels. Such lumen have tubular structures and thus projection and imaging are highly non-telecentric, containing significant perspective distortion. Given the wide range of lumen sizes inside the human body, e.g. in the gastrointestinal tract the oesophagus is 2cm in diameter while the caecum is 9cm, and the significant variation in curvature, it is challenging to develop a conventional analytical SSOP model that considers non-uniform spatial frequency projection. Thus, a data-centric approach such as GANs is promising but requires a training dataset covering many variations in lumen size and bending configuration across different disease states, which would be infeasible to generate experimentally.

Simulated image data, specifically synthetic image dataset generation from 3D modelling or rendering software (Blender, Unity and Unreal), have shown promise for successfully training AI systems for a range of applications [19, 20]. This approach is low-cost and easily adaptable to a range of situations relevant to different applications. GANs on their own can be used to produce synthetic image data, but they can be difficult to optimise and can suffer from "mode collapse" – the generator produces a limited variety of samples, ignoring corner cases and reducing diversity in the dataset [21].

Techniques such as raytracing allow real-life objects to be modelled realistically due to physically accurate lighting, shading and textures. Ray-tracing rendering engines share some similarities with Monte-Carlo simulations: individual rays of light are simulated leaving a light source, traversing through materials, being scattered, reflected or absorbed, and then returning to the camera. The Cycles raytracing renderer that comes with open-source 3D rendering package Blender simulates volume scattering inside objects using a Henyey-Greenstein Phase function, which is commonly also used in Monte-Carlo simulations of tissue [22, 23]. Using a ray-tracing engine it should therefore be possible to produce physically realistic images including scattering and absorption effects. Recently, a full ray-tracing SFDI simulation model, implemented in Blender, has been demonstrated exhibiting accurate simulation of optical scattering and absorption properties [24]. The scattering and absorption properties were plotted as a function of spatial frequency and found to agree with results simulated using diffusion approximations. This approach has the advantage of being able to realistically simulate tissue optical properties, a wide range of sample geometries, and realistic lighting conditions and camera positions. Such an approach produces highly realistic simulated image data.

Finally, using a synthetic approach it is also possible to generate near-perfect 'ground truth' images. When used in a 'supervised learning' approach, this can in principle provide superior performance to using ground truths generated by processing input data through conventional algorithms which tend to degrade quality and have limited robustness to noise. Synthetic dataset generation from 3D modelling software is a powerful up-and-coming technique with start-ups such as Parallel Domain and Zumo Labs generating synthetic image datasets for training AI systems such as self-driving cars [25, 26]. Techniques such as Domain Adaptation can allow models to be trained on simulated data and then adapted to work on real datasets, an approach which has been successfully applied to multispectral imaging data [27].

Here, we demonstrate that a deep-learning model can be trained on simulated image datasets created using our Blender model to reconstruct maps of reduced scattering and absorption for a range of realistic sample geometries and imaging set-ups. Specifically, we use a modified version of the GANPOP model (concept shown in Figure 1) because the direct production of optical property maps avoids the needs for calibration images at intermediate stages: calibrations are implicitly learned by the GAN by comparison with ground truths. We show that the model can be trained to work when imaging inside cylindrical structures, representing lumen, providing contrast for tumour-like structures. The ability of our system to generate ideal 'ground truth' images, makes it possible in principle to obtain superior performance to conventional SFDI reconstruction algorithms. The flexibility of our model also allows training a model to deal with 'edge cases,' such as high values of absorption or scattering or highly reflective materials. Finally, we show that our trained model has inherent domain adaptation ability by cross-validating on experimentally acquired SFDI data.

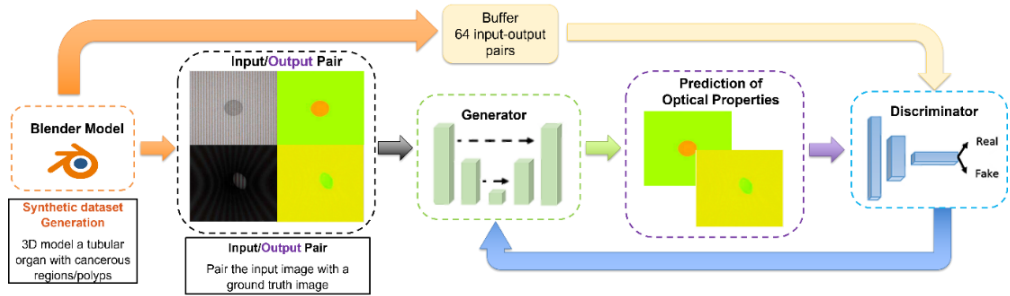

Figure 1. Proof of concept diagram for training a GANPOP with synthetic data. Blender automation generates training data which is then fed into a GAN and processed to get optical property maps.

## 2. Method

### *2.1 Blender Model*

Our simulation model is created by applying custom material properties to 3D objects in Blender. Specifically, scattering is simulated by mixing two of Blender's prebuilt shaders: the transparent bi-directional scattering distribution function (BSDF) which adds transparency to the material without adding refraction; and the subsurface scattering which simulates light rays penetrating the surface of the material and bouncing around inside until they either escape or are absorbed. A weight factor we term the 'scattering factor' is varied between 0 and 1 to control how much of the composite materials' properties arise from the subsurface scattering component vs the BSDF component. A scattering factor of 0 results in a fully transparent material whilst a scattering factor of 1 results in fully subsurface scattering material. Absorption is implemented by creating a custom material to simulate the transmissivity of light. This was implemented with a transparent BSDF shader equally mixed with a refraction BSDF shader with a refractive index of 1.43. This weighted 'absorption factor' varies the observed colour of the refraction BSDF block from 0 (black, fully absorbing) to 1 (fully transparent). The 'final factor' was a weighted mix of the absorbing and scattering components, where a factor of 0 simulates a material only dependent on the absorbing component and a factor of 1 simulates a material only dependent on the scattering component. Blender's ray-tracing engine Cycles was then used to render realistic images of the desired scene as described previously [24], with the power of the light being set to 3.5W.

Five separate Blender models were designed to create a variety of synthetic data, with each of the datasets used to test how the GANPOP model responded to different sample types and geometries. The first model was used to generate SFDI images with flat sample geometries containing only one material, which we term the 'rectangular' model. The second model generates images of flat samples containing 2 materials with different optical properties meeting along a curved boundary, which we term the 'rectangular+curved' model. The third model generates images of flat samples containing 3 materials with different optical properties that meet along a ragged boundary, which we term the 'rectangular+ragged' model. The fourth model is based on the 'rectangular+curved' model but with the addition of simulated tumours, modelled as spheroids, which we term the 'rectangular+tumour' model. The fifth model produces image of the projection of an SFDI pattern into a simulated lumen (modelled as a hollowed-out cylinder) with polyp structures (modelled as spheroids), which we term the 'cylinder+tumour' model. This last model is designed to be typical of images that would be seen through an endoscopic SFDI system deployed in a tubular organ. For convenience, we group data generated from Blender models 1-4 into a single dataset termed the 'complex rectangular' dataset, which is intended to emulate the typically more complex sample shapes encountered in experimental systems.

## 2.2. Dataset Generation

To produce a dataset for the GANPOP we must pair two sets of images of equal dimensions. The first image is the SFDI projected input image with a spatial frequency of $0.2mm^{-1}$ and the second is the co-registered ground truth optical property map for that image. To examine model generalisation across spatial frequencies, we produce validation datasets at 2 additional spatial frequencies 10% above and below this ($0.18mm^{-1}$ and $0.22mm^{-1}$). Ground truth optical property maps were generated by replacing the scattering and absorption material inputs with pure colours, unaffected by lighting and shading. Specifically, the red (R) channel represents absorption and the green (G) channel represents scattering, whilst the blue channel is held constant at zero. This approach matches that used for the original GANPOP, where a highly absorptive material appears red, and a highly scattering material appears green. R and G channels were each varied between 0.05 and 0.95 which are scaled to 8-bit integers ranging from 13-242. The display device was set to 'None' to allow for linear control of the RGB content of the image.

The 'keyframe' approach to animation supported by Blender was used to automatically vary optical and geometrical properties of samples and output large numbers of rendered images to create the dataset. For the tumour spheroid dataset, the XYZ parameter was used to control side-to-side (X), front-to-back (Y) and top to bottom (Z) translation of the spheroids whilst the scale parameter was used to change the proportions of spheroids on the sample. A value parameter block was used to set the final, absorption and scattering factors of the material as well as the colours in the ground truth optical property maps. Using the keyframe approach resulted in a smooth animation as Blender can be configured to linearly interpolate between set values. Final factor was varied between 0.05 and 0.95 resulting in the proportion of absorption to scattering to change from 0.95:0.05 to 0.05:0.95: that is, the material would change from a more absorbing material to a more scattering material. Varying a single parameter in this way maps out a 1-D path within the 2-D optical property space. To validate performance in a larger range of this space, the absorption and scattering factors are also varied.

The start and end frames are used to control the overall size of the dataset e.g. for an animation containing 100 frames, the start frame is set to 1 and the end frame to 100. When rendered with Cycles, each frame gives an accurate, spatially modulated image and at the end of the animation, Blender produces a synthetic dataset of 100 images for the above example. For speed optimisation, the spatial resolution of the RGB images is set to 256 x 256 for both the input and ground truth images. Both image datasets were simulated separately, paired in a pre-processing step and then fed to the GAN. Combined, the synthetic dataset comprises of RGB images of size 512 x 256, which could then be used for training and testing.

## 2.3. GANPOP Architecture

The GANPOP model was based on an adversarial training framework as this framework has the ability to learn a loss function during training, avoiding uncertainties commonly seen with handcrafted loss functions [28, 29]. The generator was designed as a modified U-Net consisting of an encoder and a decoder with skip connections [30], and contained properties of a ResNet, such as having short skip connections within each level [31]. U-Net architectures have also been successfully applied to produce demodulated AC and DC images from SSOP inputs [11].

The discriminator was created with a three-layer classifier with leaky ReLUs and makes classification decisions depending on the current batch as well as a randomly sampled batch of 64 previously generated image pairs. Full details of the GANPOP architecture can be found in [14].

In essence, the generator predicts pixel-wise optical properties from SFDI images while the discriminator classifies whether pairs of SFDI images and optical property maps were real or fake. Both neural networks are trained together in a game-like fashion. The generator tries to fool the discriminator by producing hyper-realistic images that match the training dataset. On the other hand, the discriminator tries not to be fooled by trying to correctly distinguish between fake and real data. The discriminator also provides feedback to the generator to help increase the accuracy of overall GAN.

## 2.4. Training

The generator and discriminator are trained iteratively, with spectral normalization used to stabilise training [32]. The objective (loss) of the GAN can be summarised as follows:

$$L_{GANPOP}(G,D) = E_{x,yp_{data}(x,y)} + E_{xp_{data}(x)}\left[D(x,G(x))^2\right], \quad (1)$$

where G is the generator, D the discriminator, and $p_{data}$ is the optimal distribution of the data. To further stabilise the GAN as well as minimise the distance from the ground truth distribution a loss term, $L_1$ was added to the GAN loss:

$$L_1(S) = E_{x,yp_{data}(x,y)}\left[||y - G(x)||_1\right]. \quad (2)$$

Hence the total loss of the GANPOP can be expressed as:

$$L(G,D) = L_{GANPOP}(G,D) + \lambda L_1(G), \quad (3)$$

where λ is the regularisation parameter, which was set to 60 for all experiments. For training, 200 epochs were used and the *Adam* optimiser was used with a learning rate of 0.0001 for half of the epochs then linearly decayed for the remaining half. Further explanations of (1) , (2) and (3) can be found in [14].

The training code was implemented using Pytorch 1.9.0 on Windows 10 and a NVIDIA GeForce GTX 980 was used for training, testing, pre-processing and post-processing. Both the discriminator and generator were initialised with a Gaussian distribution with a mean of 0 and standard deviation of 0.02.

Optical property maps were found by isolating the red and green channels of the generated output image and then applying an appropriate scaling. All output images were normalised to have a consistent representation in the 8-bit colour scale, a technique commonly used in convolutional neural networks. In previous work [14], the maximum value of 255 was defined to be 0.25 mm$^{-1}$ for $\mu_a$ and 2.5 mm$^{-1}$ for $\mu_s$'. However, here we are mapping between parameters in Blender (the various *factors*), which can be considered as proxy values for $\mu_a$ and $\mu_s$' that can be mapped back to these parameters via a calibration process such as that presented in [24].

## 2.5. Error Metric

Normalized Mean Absolute Error (NMAE) is used to evaluate the performance of the 3 GAN models trained on synthetic data and the GANPOP pretrained on experimental data. NMAE is defined as:

$$NMAE = \frac{\sum_{i=1}^{T}\left|p_i - p_{i_{ref}}\right|}{\sum_{i=1}^{T} p_{i_{ref}}}. \quad (4)$$

where $p_i$ is the predicted pixel data, $p_{i_{ref}}$ is the actual ground truth pixel data and T is the total number of pixels. A smaller NMAE indicates a better prediction by the GAN. Note that $p_{i_{ref}}$ was calculated from the simulated ground truth Blender output data for the 3 synthetic models but for the experimental model was calculated by applying standard SFDI processing algorithms to input data. To see pixel-per-pixel differences, normalized error was calculated for both the absorption and reduced scattering, then displayed as a percentage on a matplotlib

GUI. This allowed for fair visual comparisons and allowed us to see spatially where the model failed and which images it failed on:

$$p_{diff}^{i} = \frac{\left|p_i - p_{i_{ref}}\right|}{p_{i_{ref}}} \qquad (5)$$

### 2.6. Validation

Each model was validated using a subset of the generated images that were not used in training. Further, to investigate the applicability of our synthetic-data-trained model on real experimental data, a cross-validation was performed with the GANPOP model pre-trained on an experimental dataset. Validation data from the synthetic dataset was input into the model trained on experimental data and vice-versa. In each case the NMAE relative to the appropriate ground truth was computed. In the synthetic model the red and green channels of the ground truth, which can only vary from 0 to 255, are scaled relative to the absorption and final factors, whereas in the experimental model they are scaled relative to real scattering and absorption coefficients. We thus expect to observe an arbitrary scale factor between the models.

## 3. Results

### 3.1. Rectangular GANPOP

We first trained the GANPOP on the complex rectangular dataset, which comprised 50 images each from the 'rectangular', 'rectangular+curved', 'rectangular+ragged' and 'rectangular+tumour' i.e. 200 images in total. The dataset was split randomly into 140 training images and 60 validation images. The NMAE of scattering and absorption for the rectangular GANPOP are plotted in Figure 2. All validation images were encapsulated within a 2.5% absorption and reduced scattering NMAE envelope. Including anomalies, the average NMAE was calculated to be 1.2% for $\mu_a$ and 1.1% for $\mu_s$', which compares very favourably to values of 7-12% typically found in GANs trained on experimental data [10, 11].

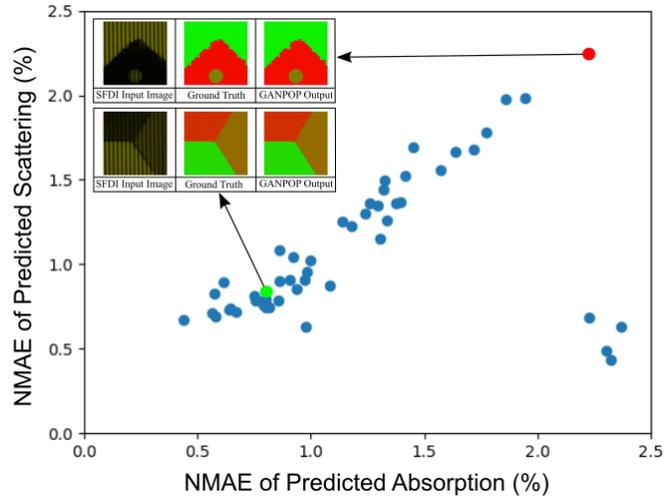

Figure 2. A scatter plot of the NMAE of scattering and absorption coefficients for the rectangular GANPOP, tested on the validation samples. Examples of a low (green) and high error result (red) are shown inset. Each result comprises a SFDI input image, a ground truth and GANPOP output.

## 3.2. Cylindrical GANPOP

Next, a GAN was trained with images created from the 'cylinder+tumour' Blender model. The 320-image dataset comprised of 200 images where a SFDI pattern was projected into a simulated lumen and 120 images where polyp structures were scattered around another lumen with varying sizes, positions and optical properties. These split randomly into 256 training and 64 validation images. The projection onto a cylindrical surface created substantial fringe distortion and thus varying spatial frequencies. However, the GAN model was still effective in recovering optical properties, with Figure 3 showing the result of testing the GAN. A notable increase in error can be seen in the NMAE envelope when compared to the previous model, with the NMAE increasing to 5.5% for absorption and 6.4% for reduced scattering. This increased envelope error may be due to the aforementioned variation in spatial frequency in addition to the reduced optical power at increasing distances further down the lumen. Nevertheless, the average NMAE was still found to be low, at around 1.0% for $\mu_a$ and 1.2% for $\mu_s'$, confirming the adaptability of the GANPOP architecture.

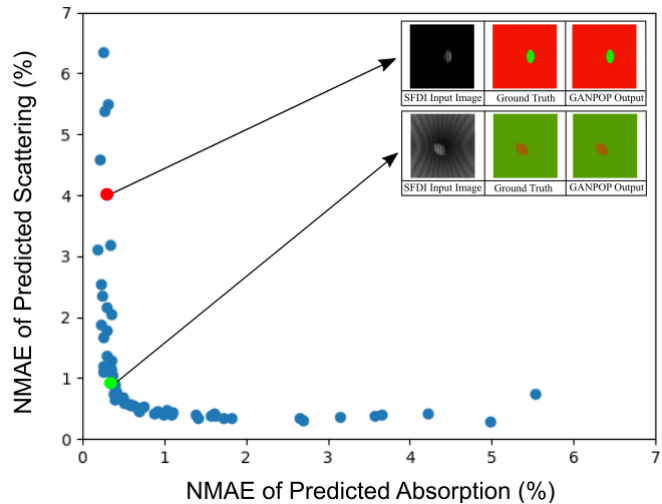

Figure 3. A scatter plot of the NMAE of reduced scattering and absorption coefficients for the cylinder GANPOP, tested on the 'cylinder+tumour' validation dataset. Examples of low (green) and high error (red) images are shown inset.

## 3.3. Training Performance

Across the four Blender models in the rectangular dataset, the average time taken to generate a SFDI image was 3.7 seconds and 1.1 seconds to generate a ground truth image. In contrast, the cylindrical dataset took, on average, 4.6 seconds to generate an SFDI image and 0.7 seconds to generate a ground truth image. For a 200-image dataset, it took 15 minutes to generate the rectangular dataset and 18 minutes to generate the cylindrical dataset.

For the rectangular GANPOP, where a training dataset of 140 images was used and the number of epochs was set to 200, the time taken to train the model was 2 hours and 30 mins. The cylindrical GANPOP with a training dataset of 256 images and 200 epochs, took approximately 4 hours to train. Testing both GANPOP models on their respective testing datasets was very quick, with the rectangular model taking 11 seconds to process 60 images and the cylindrical model taking 12 seconds to process 64 images showing the potential of using a GANPOP in clinical settings. Repeating the training process 5 times using randomized training-validation splits and starting weights shows robust convergence to acceptably low MSE values, as shown in Figure 4.

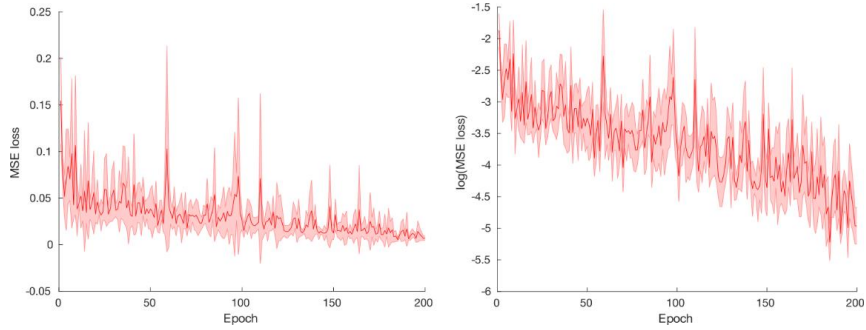

Figure 4. Plots of the MSE loss function used to train the network as a function of epochs. Shaded error bar indicates standard deviation over 5 runs using randomised starting weights and training-validation splits: A) linear scale, B) log scale

### 3.4. Spatial Characteristics of Error

Although both GANPOP models detected the simulated polyps correctly with high visual accuracy, it is insightful to analyse the main spatial sources of error. Figures 5 and 6 show optical property maps of low NMAE and high NMAE examples respectively for the cylindrical model as well as the individual R ($\mu_a$) and G ($\mu_s'$) channels. Looking at the difference maps, we can see that for the low-error result the error arises at the edges of the tumour, with some parts of the edge displaying percentage errors up to 20%. This may simply be an edge effect – light in this region can pass through multiple different material types before returning to the camera, causing increased ambiguity in recovering absorption and scattering. Despite this, for the areas surrounding the tumour the error can be seen to be near 0%. Looking more closely at the high-error result in Figure 6, much of the error comes from the area surrounding the tumour for the reduced scattering difference and the tumour itself for the absorption difference. Moreover, the error can be seen to reach 25% just below the tumour for scattering and 15% at the tumour for absorption. However, we note that many of the 'bad' examples are not representative of biologically relevant scenarios: the bulk of material in such examples is so highly absorbing as to appear black, which is not typically observed in live tissue.

We next validated the performance of the rectangular and cylindrical models using spatial frequencies 10% above and 10% below the base 0.2mm$^{-1}$ spatial frequency. A 150-image test dataset was generated using the 'rectangular+tumour' and 'cylindrical+ tumour' models. For both GANPOPs, one new validation set of 50 images was created for each of the 2 additional spatial frequencies. The validation NMAE increased for absorption coefficient (to 3.3% for +10% SF and to 3.6% for -10% SF) and for scattering (to 5.9% for +10% SF and 5.5% for -10% SF). This represents a slight increase in error compared to typical NMAE values for experimental datasets, indicating that our trained GAN is relatively robust to variation in spatial frequency and is not significantly overfitting.

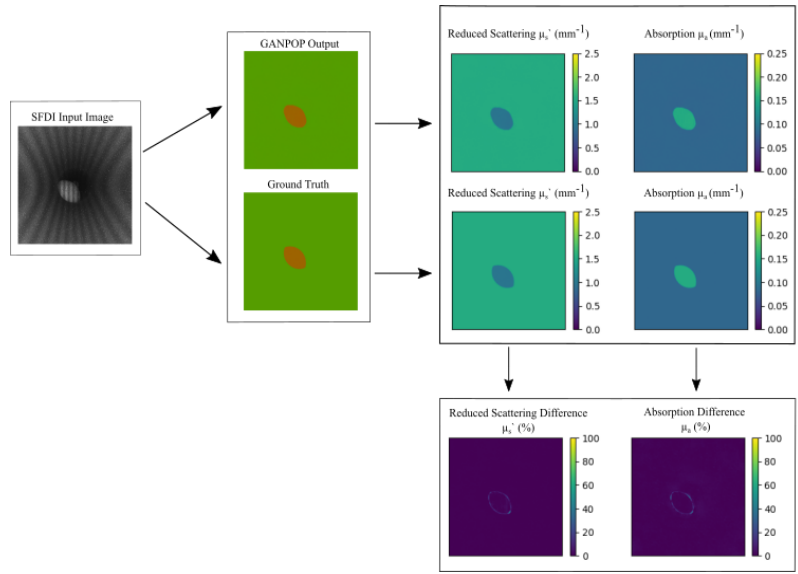

Figure 5. The Optical Property and Difference Maps of a typical low NMAE result. The SFDI input image, GANPOP Output and Ground Truth can be seen on the left. The reduced scattering, absorption and the respective difference maps can be seen on the right.

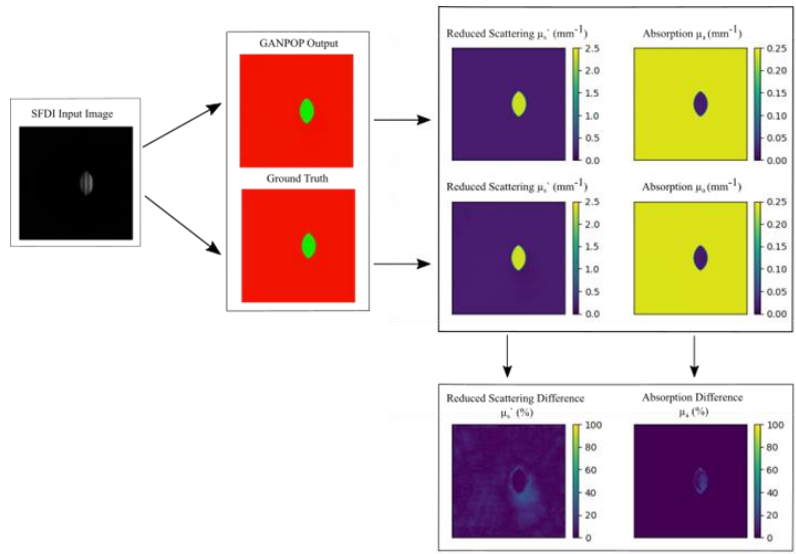

Figure 6. The Optical Property and Difference Maps of a typical high NMAE result. The SFDI input image, GANPOP Output and Ground Truth can be seen on the left. The reduced scattering, absorption and the respective difference maps can be seen on the right. For this image specifically, the absorption and scattering values were designed to be at extreme ends of the possible range.

### 3.5. Expanding optical property parameter space

In order to explore a larger part of the space of possible optical properties, different combinations and ranges of absorption, scattering and final factors are tested. This was to ensure that optical property space was adequately covered in the training data to avoid overfitting. From Table 1, it can be seen that the mean of the average absorption NMAE for the rectangular model was 1.29%, whilst the mean of the average scattering NMAE was 1.18%. On the other hand, the mean of the average of absorption and scattering NMAE for the cylindrical model was found to be 1.36% and 1.62% respectively. For both models, the larger the variation in the Blender factors the larger the NMAE was. Nevertheless, this deviation from the mean was small, demonstrating our model robustness across the wider optical property space.

| Blender Factors | | | Average absorption NMAE (%) | | Average scattering NMAE (%) | |
|---|---|---|---|---|---|---|
| Final Factor | Abs Factor | Sct Factor | Rectangular | Cylindrical | Rectangular | Cylindrical |
| 0.05-0.95 | 1.00 | 1.00 | 1.19 | 1.01 | 1.11 | 1.20 |
| 0.05-0.95 | 0.05-0.95 | 1.00 | 1.12 | 1.30 | 1.12 | 1.51 |
| 0.05-0.95 | 1.00 | 0.05-0.95 | 1.31 | 1.44 | 1.18 | 1.78 |
| 0.05-0.95 | 0.05-0.95 | 0.05-0.95 | 1.53 | 1.67 | 1.30 | 2.00 |

Table 1: A table showing the average absorption NMAE (%) and average scattering NMAE (%) for a range of different combinations of absorption, scattering and final factor Blender values for the rectangular and cylindrical GANPOPs.

### 3.6   Cross-validation with experimental data

To cross-validate our GAN with the experimentally trained GAN, we first corrected for the scale factor introduced by the different mappings between absorption and scattering coefficients and 8-bit image pixel values (0 to 255). We then tested our model trained on synthetic data on the experimental SFDI dataset. To encourage proper domain-transfer, we retrained our model with our original synthetic dataset plus a small sample (10%) of experimental images leading to a new GAN that can evaluate the optical properties of both real and synthetic data, which we term GANPOP-SR (GANPOP with synthetic and real data).

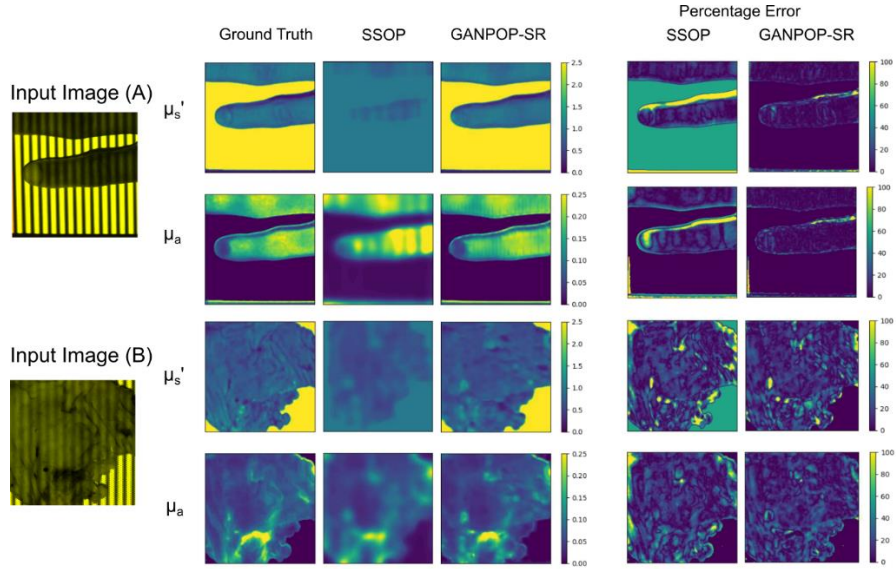

Figure 7. The optical property and difference maps of experimental sample images input into conventional SSOP and Chen's GANPOP: A) human hand, B) pig tissue.

The experimental image dataset used came from the original GANPOP system (denoted 'Chen's GANPOP' here for clarity) [10] comprising of 78 in-vivo and ex-vivo SFDI images of human oesophagi, feet and fingers, pig tissue and phantoms. Note that we removed the blue channel of the synthetic dataset to match the way the test dataset was presented in [10]. Across the entire dataset, the NMAE was found to be low, with an average value of 10.3% for absorption and 6.6% for scattering. Figure 7 shows the results of the cross validation, with Figure 7a and 7b showing examples of images where the GANPOP-SR model accurately constructed the optical property maps. For comparison, the mean SSOP NMAE of this dataset was previously computed as 23.1% for absorption and 18.7% for scattering [10].

Next, we input synthetically generated data into the model trained on experimental data. Specifically, we tested using the complex rectangular dataset which was generated from models 1-4. Across the entire dataset, the NMAE was also found to be low, with an average of 6.3% and 11.9% for absorption and reduced scattering, respectively. Figures 8a and 8b show where Chen's GANPOP accurately constructed the optical property maps. We computed the mean SSOP NMAE of this simulated dataset to be 6.6% for absorption and 22.0% for scattering, demonstrating the significant improvement offered by GANPOP.

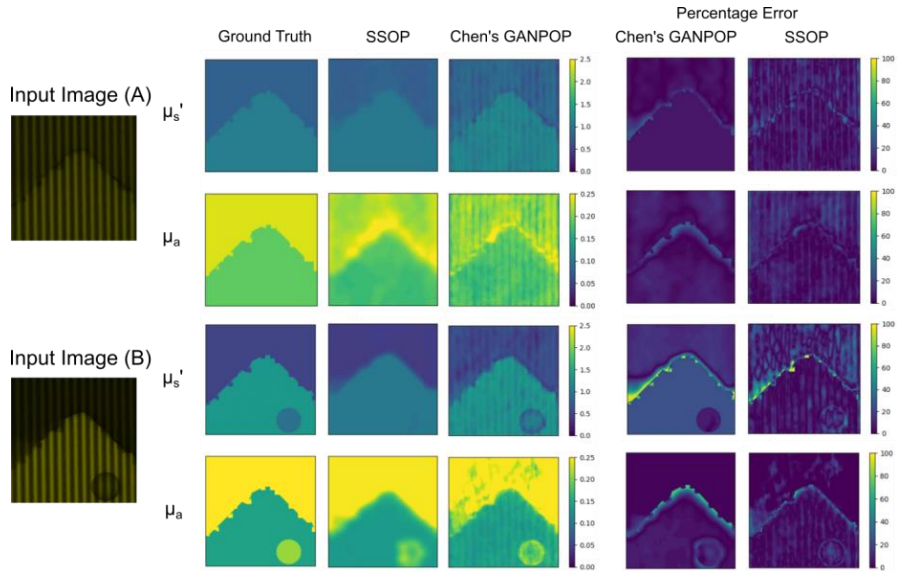

Figure 8. The optical property and difference maps of synthetic sample images input into conventional SSOP and Chen's GANPOP: A) 'rectangular+ragged' dataset sample B) 'rectangular+ragged+tumour' dataset sample.

### 4. Discussion

We demonstrate training of the Generative Adversarial Network for Prediction of Optical Properties (GANPOP) model using synthetic data from an SFDI simulation model built in open-source 3D rendering software Blender. We show that absorption and scattering can be recovered in flat and cylindrical geometries with spheroid tumours with a typical accuracy (NMAE) <1.2% and visually accurate property maps, which compares favourably to the ~10% NMAE achieved by GANPOP trained on experimental images. The high accuracy is due in part to the use of 'perfect' ground truths for supervised learning, use of larger training sets enabled by automated simulation and reduced noise compared to experimental images. However, we also show that our simulation model and simulation trained GANPOP is compatible with real-world experimental data via cross-validation. We find that when running the experimentally-trained GANPOP on our synthetic data we achieve NMAE of 6.3% in absorption and 11.9% in reduced scattering, which is comparable to the ~10% in deep-learning based models trained purely on experimental data. This also indicates the physical validity of our Blender-based synthetic generation model. For cross-validating our synthetically trained model on experimental data, we find that best performance is achieved when a small proportion of experimental data (<10%) is added to the synthetic data, which produces NMAE of 10.3% for absorption and 6.6% for reduced scattering.

The use of simulated training data for GANPOP sets confers a number of key advantages. First, it makes it easy to generate large training datasets. Using Blender's animation features it is straightforward to automatically vary many material and spatial properties of the sample and create large training datasets. Here we exploit this to generate boundaries between up to 3 materials and tumour spheroids of different size, position and optical properties in both rectangular and cylindrical geometries. This wide variety of images would be very time-consuming to achieve experimentally. This flexibility allows us to establish that optical properties can be recovered from lumen using standard SFDI illumination despite being distorted. It would be challenging to adapt conventional SSOP algorithms to work in this case of non-uniform spatial frequency and so this represents an important result. In future our system could be extended to a wider range of geometrical variation, for example by considering lumen with different longitudinal curvatures and sizes, uneven sample surfaces and different shaped tumours or polyps e.g. flat polyps [33]. Moreover, the highly realistic Cycles renderer could be exploited to create more realistic scenarios such as samples with smoothly varying optical properties, specular reflections, presence of water on surfaces, different coloured materials and different camera parameters (zoom, angle, position). Compensating for these parameters with a GAN would require a very large training dataset that could not feasibly be generated experimentally.

The second advantage is the availability of perfect ground truths. In experimental datasets, optical property maps used as ground truths for supervised learning are typically constructed from raw data using conventional SFDI approaches, such as SSOP. These approaches can produce significant reconstruction error which will in turn limit the accuracy of the GAN. With a simulation model the ground truths are perfect with accuracy determined by parameters and training efficiency of the GAN. The use of perfect ground truths also enables the GAN to implicitly learn the calibration functions that are explicitly required in standard SFDI reconstruction algorithms. Furthermore, synthetic ground truths can be created for cases where conventional SFDI algorithms would not work, e.g. highly distorted linear fringes projected down a lumen. Using a GAN trained on simulated data therefore greatly simplifies application of SFDI to non-planar geometries.

There are also limitations to this approach. Notably, there is very little noise in the training datasets. Some noise is inherently introduced by the finite number of rays and reflections and by the quantisation of light levels during rendering. Effects that would be unavoidable in experimental systems such as movement artefacts, presence of other substances and sensor noise, are not included in our simulation. However, by using either the Blender compositor or post-processing in other software, it should be possible to introduce noise and motion artefacts. By adding in such functionality, our simulation could create very large datasets of highly realistic scenarios suitable for training a high-performance GAN. Another limitation is that the Cycles renderer is a tristimulus model and so only considers optical behaviour in broadband red, green and blue colour channels. Multispectral SFDI, in which optical property maps are recovered at multiple distinct wavelengths from UV to infrared, is an increasingly important technique [3]. In future, a more advanced rendering engine may be required to simulate this, although a Blender add-on that performs full Monte-Carlo simulations has recently been demonstrated [34]. Finally, our current GAN model requires a small proportion of experimental training data to enable an acceptable degree of generalisation. In future, this proportion could be reduced further by increasing the variation within the dataset to encourage generalisation. For example, we could use a parameter randomisation approach instead of a linear parameter sweep. A greater number spatial frequencies could be used for training data although we found that our current model maintained high accuracy when validated with different spatial frequencies (+10% and -10%) suggesting a degree of inherent generalisation to spatial frequency. Other parameters such as sample to camera distance or angles may also be varied to improve generalisation.

We envisage that our tool will play a key role in designing future SFDI systems for use inside tubular lumen. For example, it could be used to design suitable illumination patterns that suffer less fringe distortion and deliver uniform spatial frequencies and illumination power in tubular organs. Additionally, domain-transfer techniques could be used to refine the trained model using experimental data but still achieve higher accuracy because of perfect ground truths. Such a model could identify subtle changes in optical properties that may not be visible using conventional SFDI reconstruction techniques but are important indicators of early disease e.g. flat lesions in the oesophagus [35]. More generally, our simulation-to-reconstruction system represents an example of 'inverse graphics' in which parameters of the generative model are inferred from a rendered image [36]. By using more data points, e.g. multiple camera views or illuminations, it may be possible to infer more information about the model including 3D shape, scattering anisotropy and surface roughness, which would be highly valuable for detecting features relevant to a wider range of diseases.

## 5. Conclusion

We demonstrate the training of a generative adversarial network using synthetic data for recovery of optical property maps in a rectangular and cylindrical image dataset with simulated spheroidal tumours. We show that the model works inside lumen like cylindrical structures and rectangular planes and can recover contrast for tumour-like structures in a variety of positions, sizes and geometries with a normalised error <1.2% for scattering and absorption recovery. Ideal 'ground truth' property maps enable superior performance to conventional SFDI reconstruction algorithms (error ~30%) and GANs trained on experimental data (error 10-15%). Moreover, we demonstrate that our model has some inherent domain adaptation ability by validating it on experimentally acquired SFDI data with error $< 12\%$. In future, this GAN could be trained on a combination of real and synthetic data, leading to a system that will better generalise to the wide range of irregular geometries encountered in real tubular organs. Resulting optical property maps could aid radiologists to detect diseases such as cancer at an early stage, e.g. during routine screening.

### Acknowledgements

The authors acknowledge support from a UKRI Future Leaders Fellowship (MR/T041951/1) and an EPSRC Studentship (2268555).

### Disclosures
The authors declare no conflicts of interest.

### Data availability statement
Data underlying the results presented in this paper are available in Ref. [37].